%% file: LHCP2014_baglio.tex
\documentclass[10pt]{article}
\usepackage{graphicx}

\def\Title#1{\begin{center} {\Large #1 } \end{center}}
\def\Author#1{\begin{center}{ \sc #1} \end{center}}
\def\Address#1{\begin{center}{ \it #1} \end{center}}

\newcommand\pubblock{\rightline{\begin{tabular}{l} Proceedings of the Second Annual LHCP\\ \pubnumber\\
         \pubdate  \end{tabular}}}

\newenvironment{Abstract}{\begin{quotation} \begin{center} 
             \large ABSTRACT \end{center}\bigskip 
      \begin{center}\begin{large}}{\end{large}\end{center} \end{quotation}}

\newenvironment{Presented}{\begin{quotation} \begin{center} 
             PRESENTED AT\end{center}\bigskip 
      \begin{center}\begin{large}}{\end{large}\end{center} \end{quotation}}

\input econfmacros.tex

\usepackage{color}
\usepackage{hyperref}

\newcommand\pubnumber{KA-TP-24-2014; SFB/CPP-14-65}

\newcommand\pubdate{\today}

\def\affiliation{
Institut f\"{u}r Theoretische Physik,\\
Karlsruhe Institute of Technology (KIT)\\
Wolfgang-Gaede Strasse 1, Karlsruhe D-76131, Germany}

\def\support{\footnote{Work supported by the DFG under the SFB TR-9
    Computational Particle Physics}}

\begin{document}

\large
\begin{titlepage}
\pubblock

\vfill
\Title{A theoretical status of the triple Higgs coupling studies at
  the LHC}
\vfill

\Author{Julien Baglio \support}
\Address{\affiliation}
\vfill
\begin{Abstract}
Now that a Higgs boson has been discovered at the LHC, measuring its
couplings to other particles is the next important step. In order to
probe the electroweak symmetry breaking mechanism at its core it is
crucial to reconstruct the scalar potential and hence measure the
triple Higgs coupling at the LHC. We present a review of the main
Standard Model Higgs boson pair production mechanisms in which the
triple Higgs coupling plays a role and present the latest
phenomenological analyses in view of a high luminosity LHC. One
example of an analysis in the Two-Higgs-Doublet model will also be
given as an illustration of an extended Higgs sector.
\end{Abstract}
\vfill

\begin{Presented}
The Second Annual Conference\\
 on Large Hadron Collider Physics \\
Columbia University, New York, U.S.A \\ 
June 2-7, 2014
\end{Presented}
\vfill
\end{titlepage}
\def\thefootnote{\fnsymbol{footnote}}
\setcounter{footnote}{0}
%

\normalsize 


\section{Introduction}

After the discovery in 2012 at CERN~\cite{Aad:2012tfa} of a Higgs
boson~\cite{Englert:1964et} it is of utmost importance to pin down its
properties, notably through couplings measurements. It looks like a
Standard Model (SM) Higgs boson so far~\cite{ATLAS:conf} but there is
still the possibility of a beyond-the-SM (BSM) interpretation of the
data.

After electroweak symmetry breaking (EWSB), the scalar potential
contains triple and quartic Higgs couplings. Their measurements would
allow for the reconstruction of the scalar potential. It has been
shown that the quartic Higgs coupling is not accessible at current of
foreseen collider energies of order 100 TeV~\cite{Plehn:2005nk}. This
is the reason why the focus is on the triple Higgs coupling that is
accessible through Higgs boson pair production. It has been the focus
of early theoretical studies at
leptonic~\cite{Boudjema:1995cb,Djouadi:1999gv} and
hadronic~\cite{Djouadi:1999rca} colliders and a detailed analysis of
the $b\bar{b}\gamma\gamma$ search channel in the early 2000s,
including a fit to the $m_{HH}$ distributions, has stated that
excluding a vanishing triple Higgs coupling would be possible at the
LHC with a very high luminosity of 6 ab$^{-1}$~\cite{Baur:2002rb}.

This review deals with the recent theoretical calculations of the SM
production mechanisms and the state-of-the-art phenomenological
analyses. Numerous BSM studies have also been performed and one case
example will be given in the context of the Two-Higgs-Doublet model
(2HDM) of type II.

\section{SM Higgs boson pair production at the LHC}

\subsection{Overview of the main channels}

\begin{figure}[htb]
\centering
\includegraphics[scale=0.5]{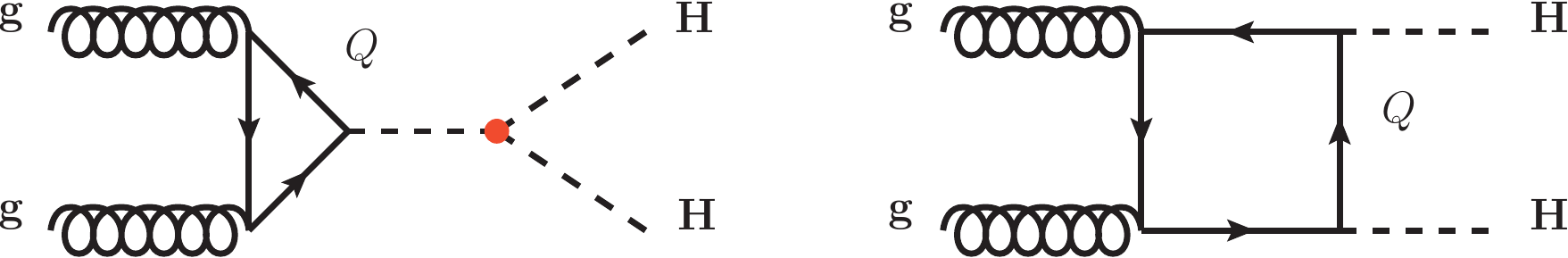}

\vspace{4mm}\includegraphics[scale=0.5]{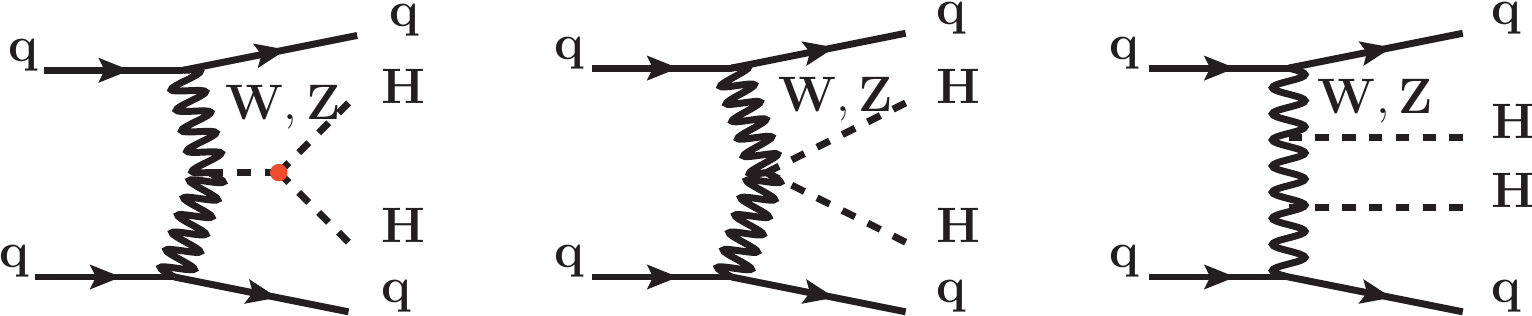}
\vspace{-2mm}\caption{Generic Feynman diagrams contribution to gluon
  fusion Higgs pair production (up) and VBF production (down). The
  triple Higgs coupling is highlighted in red.}
\label{fig:diagram}
\end{figure}

\begin{figure}
\begin{center}
\begin{minipage}[c]{5cm}
\hspace{-6.5mm}\includegraphics[scale=0.49]{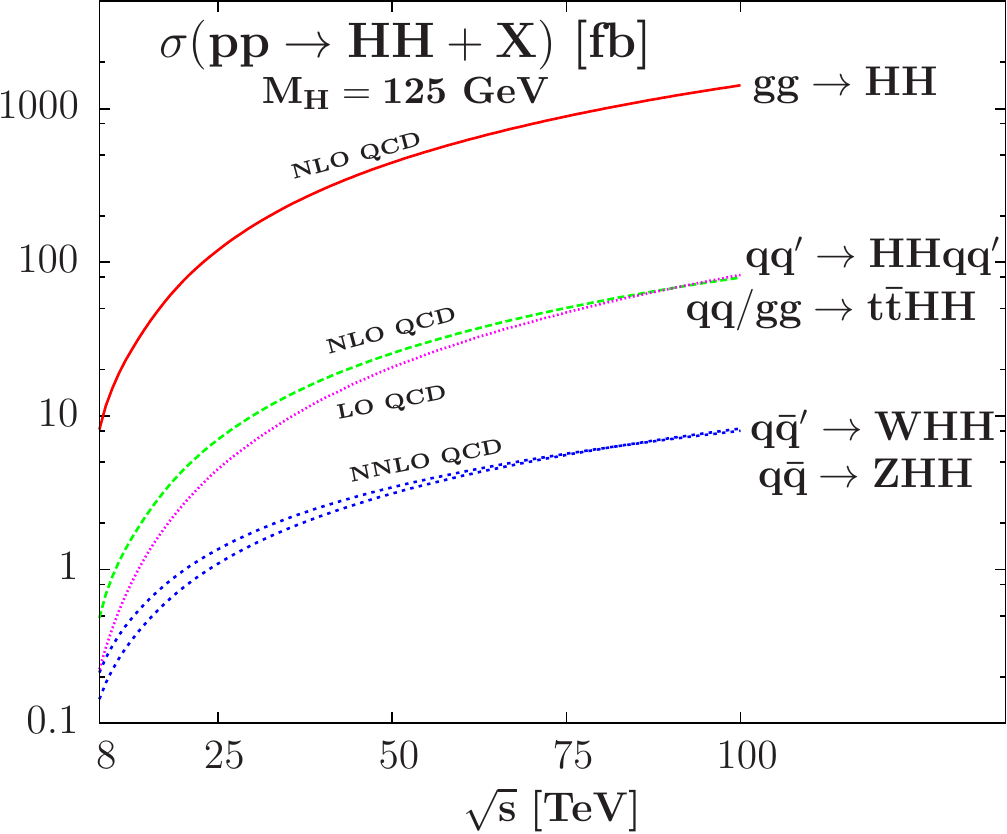}
\end{minipage}
\begin{minipage}[c]{5cm}
\vspace{-3mm}\hspace{-5.5mm}\includegraphics[scale=0.37]{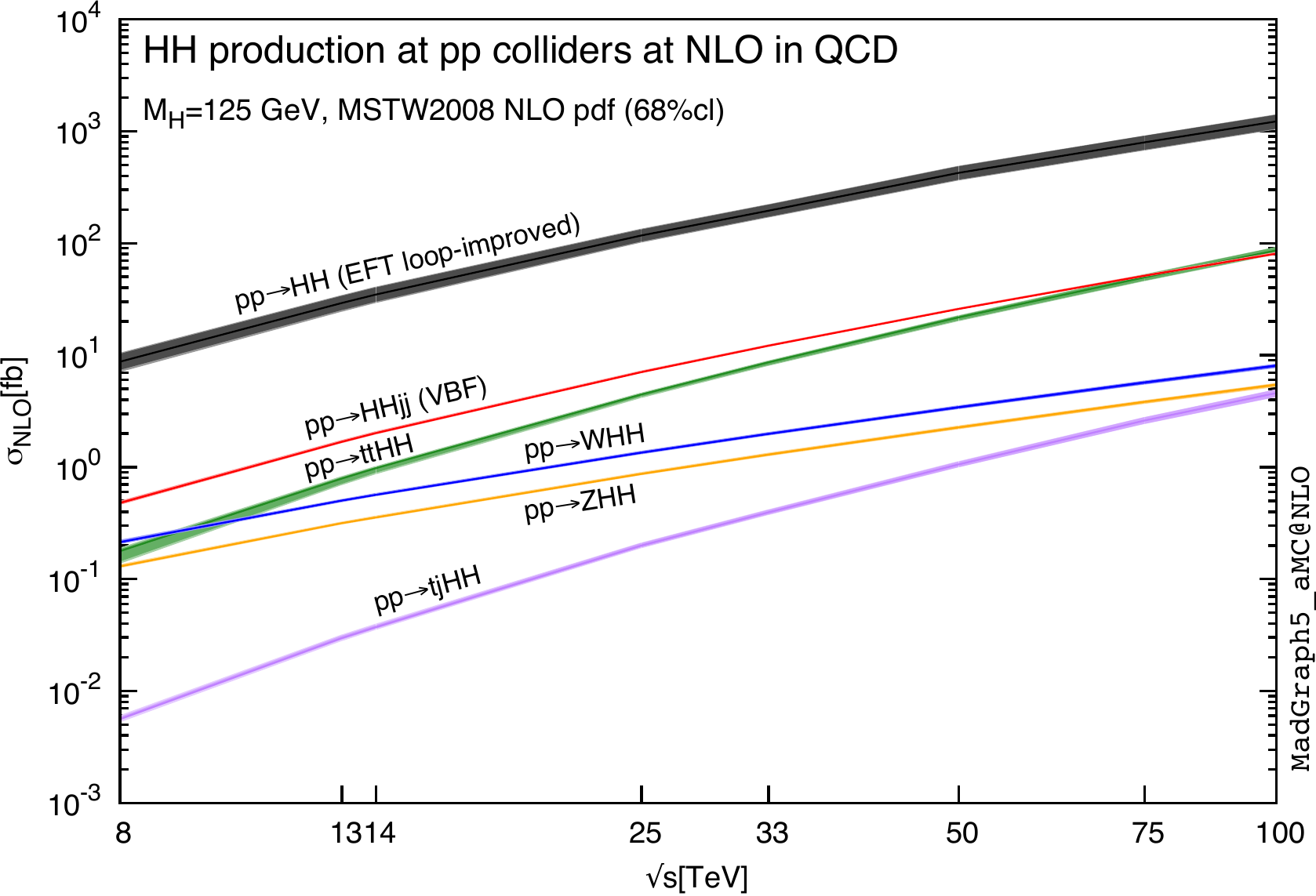}
\end{minipage}
\begin{minipage}[c]{5cm}
\hspace{5.5mm}\includegraphics[width=5.24205cm,height=4.19265cm]{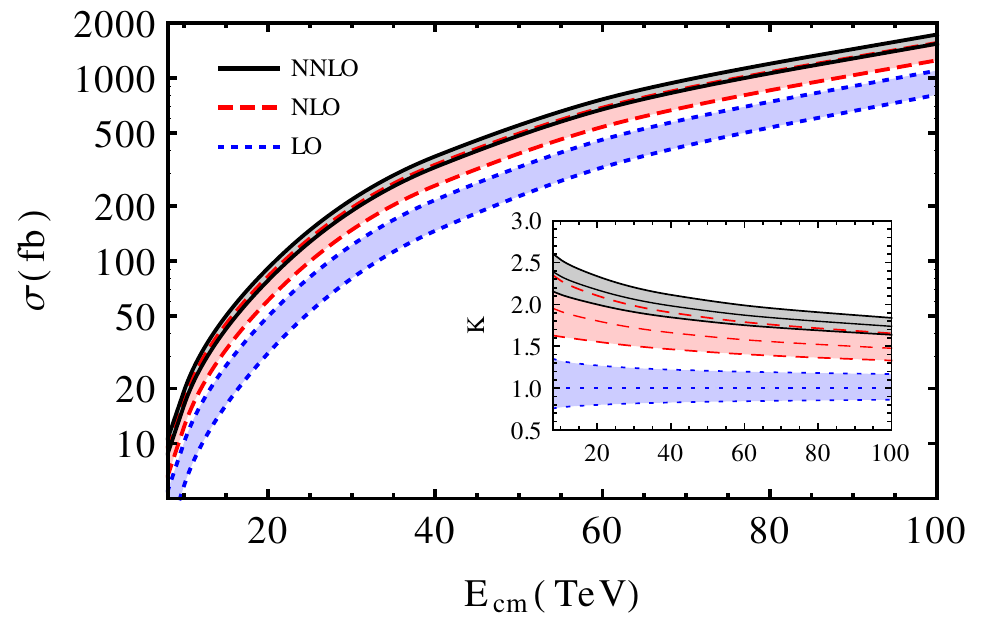}
\end{minipage}
\end{center}
\vspace{-5mm}\caption{Left and center: the total hadronic cross
  section of the main production channels of a Higgs boson pair $HH$
  (in fb) as a function of the center-of-mass-energy (in TeV); taken
  from   Ref.~\cite{Baglio:2012np} (left) and
  Ref.~\cite{Frederix:2014hta} (center). Right: the same but with
  gluon fusion only, at the NNLO accuracy and including the scale
  uncertainty, taken from Ref.~\cite{deFlorian:2013uza}.}
\label{fig:main}
\end{figure}

The main production channels for a Higgs boson pair follow the same
pattern as for single Higgs production and generic Feynman diagrams
are depicted in Fig.~\ref{fig:diagram}. In Fig.~\ref{fig:main} the
total cross section is presented as a function of the center-of-mass
energy. All cross sections are $\sim 1000$ times smaller that their
single Higgs production counterparts: a high luminosity is required to
measure the production of a Higgs boson pair.

The gluon fusion mechanism is the largest production channel. It
is mediated by loops of heavy quarks (top and bottom in the SM), see
Fig.~\ref{fig:diagram} (up). The leading order (LO) cross section
was calculated decades ago~\cite{Eboli:1987dy,Glover:1987nx} and the
process has been known for long at next-to-leading order (NLO) in QCD
in an effective field theory (EFT) approach using
the infinite top quark mass approximation~\cite{Dawson:1998py}. The
NLO $K$-factor is of the order of 2, similar to the single Higgs
production case. The major improvement in 2013 came from the extension
of this calculation up to the next-to-next-to-leading order (NNLO),
providing a $+20\%$ increase of the total cross
section~\cite{deFlorian:2013uza}, see Fig.~\ref{fig:main} (right). An
improved NLO calculation is also available~\cite{Frederix:2014hta}
including the exact real emission. A next-to-next-to-leading
logarithmic (NNLL) resummation was performed in
Ref.~\cite{Shao:2013bz} and increases the NLO cross section by $20\%$
to $30\%$, stabilizing also the scale dependence of the result. The
merging to parton showering effects for gluon fusion plus one jet has
been done in 2014~\cite{Maierhofer:2013sha} leading to a sizable
reduction of the uncertainties on the efficiencies of the cuts down to
the level of $10\%$.

The second production channel at the LHC is the vector boson fusion
(VBF). The structure of this process is very similar to the single Higgs
production case and proceeds at LO via the generic Feynman diagrams
depicted in Fig.~\ref{fig:diagram} (down). The LO cross section has
been known for a while~\cite{Eboli:1987dy,Keung:1987nw} and recently
the NLO QCD corrections have been calculated for the total cross
section and the differential
distributions~\cite{Frederix:2014hta,Baglio:2012np} and they increase
the LO result by $\simeq 7\%$. The calculation has been implemented in
the public code {\tt VBFNLO}~\cite{Arnold:2008rz}. The approximate
NNLO QCD corrections have been obtained using the structure function 
approach which gives quite good results for the total cross section
and they increase the NLO result by less than
1\%~\cite{Liu-Sheng:2014gxa}.

The two other channels are of less importance, the double
Higgs-strahlung known up to NNLO in QCD~\cite{Baglio:2012np} and the
associated production with a top-antitop pair known up to NLO in
QCD~\cite{Frederix:2014hta}. A NLO interface to parton shower for
these processes as well as the first two presented has been performed
in Ref.~\cite{Frederix:2014hta}, allowing for NLO differential
predictions in all channels.

\subsection{Theoretical uncertainties on the total rates}

The gluon fusion channel is affected by sizable uncertainties of three
different types: {\it a)} the scale uncertainty  due to the
variation of the renormalization scale $\mu_R$ and the factorization
scale $\mu_F$ around a central scale $\mu_0=M_{HH}$. This provides a
rough estimate of the missing higher-order 
terms and amounts to $\simeq \pm 8\%$ at NNLO at 14
TeV~\cite{deFlorian:2013uza}, see Fig.~\ref{fig:main} (right); {\it
  b)} the uncertainty related to the parton distribution function
(PDF) and the experimental value of $\alpha_s(M_Z^2)$. This uncertainty
calculated at NLO within the MSTW2008 PDF set~\cite{Martin:2009} at
90\% CL is $\pm 7\%$ at 14 TeV~\cite{Baglio:2012np}; {\it c)} the
uncertainty related to the EFT approach (see Ref.~\cite{Baglio:2012np}
for more details), estimated to be of the order of
10\%~\cite{Baglio:2012np} and confirmed by the top mass expansion
calculation of Ref.~\cite{Grigo:2013rya}. The total uncertainty
amounts to $\pm 37\%$ at 14 TeV at NLO~\cite{Baglio:2012np}, which can
be reduced down to $\pm 30\%$ using the latest NNLO result.

The VBF channel is a rather clean process and the theoretical
uncertainties are rather small. The scale uncertainty, calculated with a
variation of $\mu_R$ and $\mu_F$ around the central scale
$\mu_0=Q^*_{W/Z}$ is roughly $ \pm 3\%$ at
NLO~\cite{Baglio:2012np}. The PDF uncertainty is limited and amounts
to $\simeq +7\% / - 4\%$ at 14 TeV. There is no EFT uncertainty and
the total theory error is $\simeq +8\% / -5\%$ at 14
TeV~\cite{Baglio:2012np}.

\section{Parton level analysis}

The Higgs pair production process needs to be measured in order to
extract the triple Higgs coupling $\lambda_{HHH}$. The total rates
being quite small, it is required in the parton level
analyses that at least one Higgs boson decays in a 
$b\bar{b}$ pair because this channel has the highest branching
fraction. There are then two main interesting final states: {\it a)}
$b\bar{b} \tau\tau$; {\it b)} $b\bar{b} \gamma\gamma$, 
rather clean but the rates are very small and there is a lot of fake
photon identification. These channels are currently used by the
experimental collaborations in their projections for the
future~\cite{ATLAS:2013hta}. All the analyses are based on the gluon
fusion production channel at 14 TeV using LO $gg\to HH$ matrix
elements normalized to the NLO total cross section and boosted
topology cuts in addition to standard acceptance cuts. The channel
$HH+2j$, including VBF production, has started to be
investigated~\cite{Dolan:2013rja}.

\subsection{\boldmath The $b\bar{b}\tau\tau$ and
  $b\bar{b}\gamma\gamma$ channels}

The $b\bar{b}\tau\tau$ channel is rather promising. When using a
$\tau$ reconstruction efficiency of 80\%, $M_{HH}> 350$ GeV and
$p_T(H)>100$ GeV as boosted topology cuts and an optimistic mass
window $112.5$ GeV $ < M_{\tau\tau} < 137.5$ GeV, this results in a
significance $S/\sqrt{B}=2.97$ already at 300 fb$^{-1}$
and 9.37 at 3 ab$^{-1}$~\cite{Baglio:2012np}, corresponding to 33 and
330 signal events respectively.

The main improvement in 2012 came from the use of the jet
substructure analysis presented in Ref~\cite{Butterworth:2008iy}. This
technique has been applied in Ref.~\cite{Dolan:2012rv} in addition to
the other cuts presented above, obtaining a signal-over-background
ratio $S/B\simeq 0.5$ and 95 signal events at 1000 fb$^{-1}$. Adding
one jet in the final state enhances the significance and $S/B\simeq
1.5$, and with additional improvements a 60\% accuracy on
$\lambda_{HHH}$ could be reached at 3
ab$^{-1}$~\cite{Barr:2013tda}. This very promising channel hence needs
a dedicated analysis by the experimental collaborations to assess the
potential difficulties of a realistic experimental environment.

The $b\bar{b}\gamma\gamma$ channel is a clean channel but rather
difficult because of the smallness of the signal rates and the large
amount of fake photons. Nevertheless it has been found in
Ref.~\cite{Baglio:2012np} that the significance could be
$S/\sqrt{B}=6.46$ at 3 ab$^{-1}$ with 47 signal events when assuming a
$b$-tagging efficiency of 70\% and including a simulation of the fake
photons. This simulation also uses the same
boosted topology cuts as above with $|\eta_{H}|<2$ and
an isolation $\Delta R(b,b)<2.5$ in addition. This promising channel
has also been part of a high energy LHC analysis~\cite{Yao:2013ika}.

Using a multivariate analysis could improve the results. It has been
found in Ref.~\cite{Barger:2013jfa} that it increases the significance
of the signal and would lead to a probe of the triple Higgs coupling
at the level of $40\%$ uncertainty at the LHC at 14 TeV using 3
ab$^{-1}$ of data.

\subsection{More improvements}

Additional improvements could increase the sensitivity of
the previous searches. In Ref.~\cite{Goertz:2013kp} it has been
advocated to use the ratios $C_{HH}$ of
double Higgs production to single Higgs production cross
sections to significantly reduce the theoretical uncertainties, down
to $\Delta^{\mu} C_{HH} \simeq \pm 2\%$ and $\Delta^{\rm PDF}C_{HH}
\simeq \pm 2\%$. This is due to the similar structure in the
higher-order corrections in both channels.

The semi-leptonic $b\bar{b}W^+W^-$ channel at 14 TeV could be also a
valuable channel when using a multivariate analysis with a possible 
significance $S/\sqrt{S+B} = 2.4$ at 600 fb$^{-1}$ already with 9
signal events~\cite{Papaefstathiou:2012qe}. In addition, the 4$b$
channel had been thought for long not to be a useful channel,
nevertheless it has been reanalyzed recently with a jet substructure
analysis and a side-band analysis and found to be interesting with 3
ab$^{-1}$ of data to constrain $\lambda_{HHH} < 1.2 \times
\lambda_{HHH}^{\rm SM}$  at 95\% CL~\cite{deLima:2014dta}. More
experimental analyses are obviously required to confirm this promising
result.

\section{The 2HDM of type II: a case example of a BSM analysis}

The study of the triple Higgs coupling in various extensions of the SM
has been very active in the past few years. There are many examples,
such as with a strong Higgs sector~\cite{Grober:2010yv}, Minimal
Supersymmetric SM (MSSM) analyses~\cite{Cao:2013si}, Next-to-MSSM
analyses~\cite{Ellwanger:2013ova}, etc. As a case example we choose
here to present the CP-conserving 2HDM in which there are two Higgs
doublets leading to five Higgs bosons: 2 CP-even bosons $h$ and $H$,
one CP-odd boson $A$ and two charged Higgs bosons $H^\pm$. In the type
II version one doublet couples to the up-type fermions while the other
couples to the down-type fermions.

\begin{figure}
\begin{center}
\begin{minipage}[c]{5cm}
\hspace{-4mm}\includegraphics[scale=0.8]{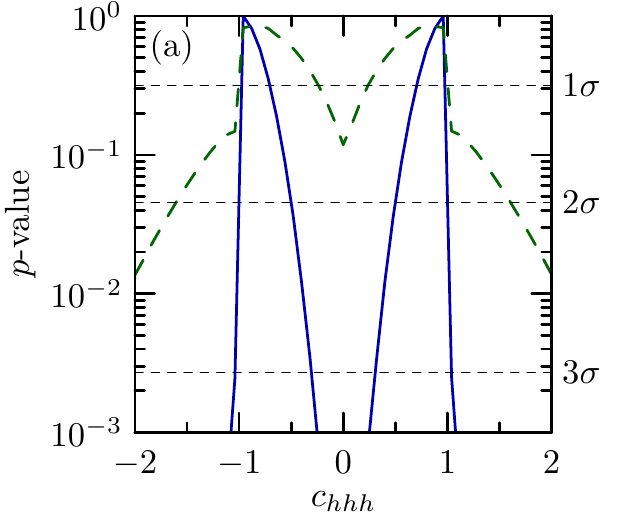}
\end{minipage}
\begin{minipage}[c]{5cm}
\includegraphics[scale=0.8]{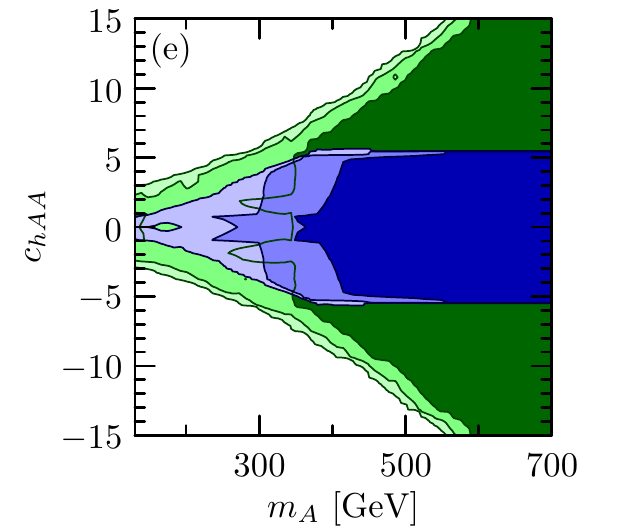}
\end{minipage}
\begin{minipage}[c]{5cm}
\hspace{3mm}\includegraphics[scale=0.5]{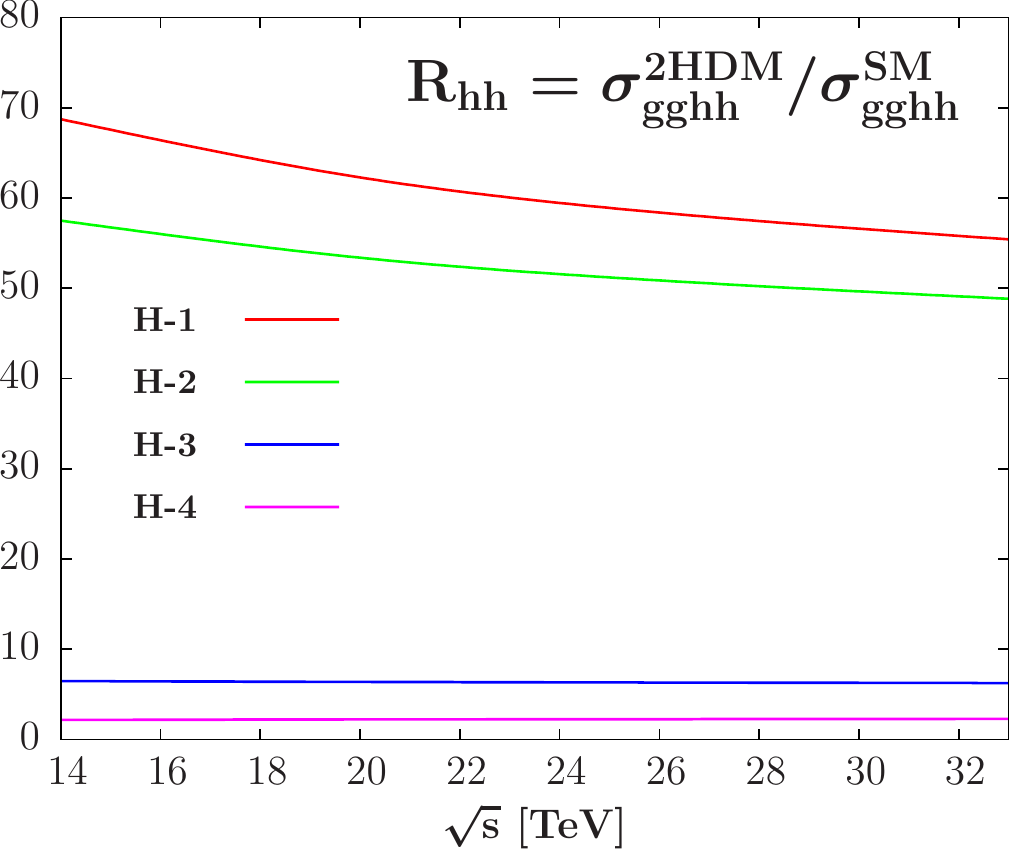}
\end{minipage}
\end{center}
\vspace{-5mm}\caption{Left: the $p$-value of the type II 2HDM with a
  fixed 2HDM/SM ratio $c_{hhh}$ for $\lambda_{hhh}$ (solid blue: tight
  perturbativity bound; dashed green: looser bound). Center: allowed
  range for the ratio $c_{hAA}=\lambda_{hAA}/\lambda_{HHH}^{\rm SM}$,
  from dark plain to light plain are the $1\sigma, 2\sigma$ and
  $3\sigma$ limits. Right: the ratio $\sigma^{\rm 2HDM}(gg\to
  hh)/\sigma^{\rm SM}(gg\to hh)$ as a function of the center-of-mass
  energy in TeV, in four benchmarks scenarios. All figures from
  Ref.~\cite{Baglio:2014nea}.}
\label{fig:2hdm}
\end{figure}

A fit using the latest experimental data as well as theoretical
constraints, especially tight perturbativity limits, has been
presented in Ref.~\cite{Baglio:2014nea}. The triple Higgs coupling of
the light $h$ cannot be enhanced compared to the standard
$\lambda_{HHH}^{\rm SM}$ (see Fig.~\ref{fig:2hdm} left). Still the triple
Higgs couplings between non-standard Higgs bosons can reach $5\times 
\lambda_{HHH}^{\rm SM}$ at $2\sigma$ as exemplified with the case of
$\lambda_{hAA}$ in Fig.~\ref{fig:2hdm} (center). Thanks to the effect of a
possible resonant heavier CP-even Higgs boson $H$ it is also possible
to greatly enhance $\sigma(gg\to hh)$ and it could be one detection
mode for the heavier Higgs boson $H$ (see Fig.~\ref{fig:2hdm}
right). These effects have also been studied in
Ref.~\cite{Moretti:2004wa}.

\section{Outlook}

Extracting the triple Higgs coupling and hence measuring the
production of a Higgs boson pair is one of the goals of the high
luminosity run of the 14 TeV LHC. Great improvements have been made in
the calculation of the SM cross sections, now reaching at least the NLO QCD
accuracy if not that of the NNLO. The theoretical uncertainty is
then reduced down to the level of 30\% in the gluon fusion
channel and below 10\% in the other channels. It is expected that the
next coming years will bring major improvements towards a fully
differential NLO calculation of the gluon fusion channel including the
full quark mass dependance. The parton level analyses, in particular
in the $b\bar{b}\tau\tau$ and $b\bar{b}\gamma\gamma$ channels, have
seen good prospects already at 300 fb$^{-1}$ and mostly at 3
ab$^{-1}$. These two channels are now under consideration by ATLAS and
CMS.

There has been also a lot of BSM activity in order to pin down
potential large effects on the triple Higgs coupling. One example is
the 2HDM of type II in which non-standard triple Higgs couplings can
reach five times the size of the standard triple Higgs coupling. Light
$hh$ pair production can also be greatly enhanced due to a
resonant heavier Higgs boson $H$. Lots remain to be done given
the vast landscape of BSM physics.


\end{document}

%% file: econfmacros.tex
\def\beq{\begin{equation}}
\def\eeq#1{\label{#1}\end{equation}}
\def\eeqn{\end{equation}}

\def\beqa{\begin{eqnarray}}
\def\eeqa#1{\label{#1}\end{eqnarray}}
\def\eeqan{\end{eqnarray}}

\let\bar=\overbar

\def\Dslash{\not{\hbox{\kern-4pt $D$}}}
\def\dslash{\not{\hbox{\kern-2pt $\del$}}}

\def\msb{{\bar{\ssstyle M \kern -1pt S}}}

